\documentclass[a4paper]{article}
\title{Parametric Equations of the Theory of Formation of
Spherical Micelles}
\author{Denis S. Grebenkov}
\date{2001}

\newcommand \ve  \varepsilon

\begin{document}

\begin{center}
\Large
{\bf Parametric Equations of the Theory of Formation of
Spherical Micelles}

\vskip 3mm

D.S.Grebenkov\footnote{ E-mail:  Denis.Grebenkov@polytechnique.fr }

\vskip 3mm

\normalsize

St. Petersburg State University, Department of Statistical Physics

ul. Ul'yanovskaya 1, Petrodvorets, 198904, Russia

\vskip 2mm

Ecole Polytechnique, Laboratoire de Physique de la Mati\`ere Condens\'ee

91128 Palaiseau Cedex, France

\end{center}

\vskip 10mm
\section*{ Abstract }

  Using the notion of aggregation work, we construct a system 
of differential \hskip 1mm equations for the aggregation number 
of micelles which is a function of the parameters of micellization
({\it parametric equations}). There are explicit solutions for
two important models of spherical micelles. Based on these solutions, 
we obtain an analytical expression for the equilibrium concentration 
of surfactant monomers and consequently for the whole spectrum of 
equilibrium concentrations of molecular aggregates in this framework. 
Accuracy of these expressions is discussed, and they are applied on 
an example of micelles formed by sodium dodecyl sulfate.

\vskip 10mm
\section*{ Introduction }

  The process of micellization is an important and interesting problem,
whose mechanism is not yet understood. Its complexity does not allow 
to describe it completely.
  If the concentration $n_1$ of surfactant monomers (amphiphiles) exceeds 
the critical micellization concentration, the monomers can form aggregates 
called {\it micelles}. There are different structures of micelles 
(spherical, cylindrical, disk-like, inverse, etc., see \cite{Israel2}), 
and they are classified by $n_1$. For a relatively small concentration $n_1$, 
the spherical structure is the most favorable\footnote{ Here we will not 
consider the supplemental conditions which can be imposed by packing constraints,
see \cite{Israel2}. }. There are two essential advantages in this case. 
First, the concentration of monomers is small and therefore one can apply 
some useful approximations. Second, the geometry of micelles is described 
by the only parameter -- the radius of the sphere. Moreover, this radius 
can be expressed in terms of aggregation number  (the number of monomers 
which form an aggregate). Consequently, it is natural to use methods of 
the nucleation theory and, particularly, the notion of aggregation work. 
One can find its description and applications in \cite{Mic1}-\cite{Mic5} 
that investigate thermodynamic properties of micelles.

\vskip 1mm

  We propose a method which allows to determine the dependence of
the equilibrium concentration of monomers and micelles on the 
essential parameters defining the process of micellization. 
We are going to establish {\it parametric equations}
which can be solved explicitly for the different model expressions
of the aggregation work. Therefore, we will able to determine 
equilibrium concentration of monomers as a function of external
parameters (e.g., temperature) in the framework of a given model. 
Consequently, it will allow to relate different models of spherical micelles 
among themselves and to compare them with regard to the experimental data.

\vskip 1mm

  The first part is devoted to introduction of the basic notions and
the essential assumptions which will be used to establish 
parametric equations in the second part. The third and the fourth parts
give solutions of these equations for the drop model and Grinin's model
of spherical micelles. The final part describes some consequences and
the possible generalizations of this formalism.

\vskip 10mm
\section{ Basic notions and essential assumptions }

  Nucleation theory is based upon the notion of {\it aggregation
work} $F_\nu $, i.e. minimal work required to form an aggregate
from $\nu $ surfactant monomers (amphiphiles), where $\nu $ is 
referred to {\it aggregation number}. Function $F_\nu $ essentially 
defines the dynamics of the system, and completely determines equilibrium 
distribution of aggregates $n_\nu $ by virtue of the aggregation number 
$\nu $ according to Boltzmann's law,

\begin{equation}      \label{Boltz}
n_\nu =n_1e^{-F_\nu },
\end{equation}
where $n_1$ is the concentration of monomers\footnote{ 
We shall use dimensionless quantities. Particularly, 
all the concentrations are expressed in convenient units: 
for example, in units of the critical micellization concentration;
the aggregation work is measured in units of $kT$,
where $k$ is Boltzmann constant, $T$ is absolute temperature.}.
Note that $F_1=0$ because the monomers already exist. 
Function $F_\nu $ depends on the aggregation number $\nu $, 
concentration of monomers in solution $n_1$, and parameters defining the 
micellization (e.g. temperature, molecular properties of solvent,
structure of monomers, etc) which are denoted as $a_1$...$a_K$.
  Note that the aggregation work $F_\nu $ does not depend on
the concentrations $n_\nu $ ($\nu >1$). Indeed, we are working
with spherical micelles, thus the concentration of monomers is
relatively small and highly dominates the concentrations of dimers, 
trimers, etc. Hence, we can assume that aggregates are formed
due to gradual attachment/detachment of monomers.
In other words, we neglect the possibilities of attachment/detachment
of dimers, trimers, etc. to the aggregate.
  Moreover, for the small concentration (i.e. the dilute solution)
the dependence of function $F_\nu $ on $n_1$ is simply\footnote{ 
This is a direct consequence of the differential form
of $F_\nu $ as difference of chemical potentials
$$dF_\nu =(\mu _\nu -\mu _0)d\nu ,$$
where $\mu _\nu $ is the chemical potential of any surfactant
``inside'' the aggregate consisting of $\nu $ molecules. Obviously,
$\mu _\nu $ is independent of the concentration of monomers in the solution
and depends only on the internal structure of aggregate. $\mu _0$ is
the chemical potential of any surfactant in the solution.
For the dilute solutions there is a simple expression for $\mu _0$
$$\mu _0=\ln n_1+\chi (T) ,$$
where $\chi (T)$ is a certain function of temperature (see \cite{Landau}).}

\begin{equation}   \label{F}
F_\nu =G_\nu -(\nu -1)\ln n_1 ,
\end{equation}
where new function $G_\nu $ does not depend at all on the concentrations
and is defined only by the parameters $a_1$..$a_K$ and $\nu $.
The condition $F_1=0$ implies $G_1=0$.

\vskip 1mm

  We make an important assumption of the {\it closure} of the system,
i.e. the total number of amphiphiles (in monomers and aggregates) $N$ is 
fixed. This is an essential condition because one can imagine some processes 
where this number changes. From now on, $N$ will be a fixed external parameter. 
In this case, the law of mass conservation can be written as

\begin{equation}   \label{masse}
N=\sum\limits _{\nu =1}^{\infty }\nu n_\nu .
\end{equation}
Note that in real physical problems the summation is usually 
limited by a certain value $\nu _m$. However, it is convenient to use
this notation, remembering that the finite sum can be ``complemented
by zeros'' (which is equivalent to rapid growth of $F_\nu $
with increase of $\nu $). For the convergence of series it is
sufficient that $F_\nu $ tends to infinity when $\nu \to \infty $, 
or that $G_\nu $ increases faster than a linear function of $\nu $.

\vskip 1mm

  As the mechanism of micelle formation is very complex, it is 
reasonable to use model expressions for $G_\nu $ which should represent
the essential properties of micellization. We represent function
$G_\nu $ as a sum of the fractional powers\footnote{ This representation
is quite reasonable: for example, in the problem of condensation
of vapor the aggregation work is the sum of the powers of $\nu ^{1/3}$,
in Grinin's model of spherical micelles -- the sum of the powers
of $\nu ^{1/2}$.} of $\nu $,

\begin{equation}   \label{G}
G_\nu =\sum\limits _{m=0}^Mc_m\nu ^{\rho m},
\end{equation}
where $\rho >0$ is a constant for the particular model\footnote{ As we use 
the model expressions, we can think of $\rho $ as a rational number. 
If the sum (\ref{G}) includes the linear term, we can take $\rho =1/m_0$ with 
a natural number $m_0$.}, the coefficients $c_m$ of this decomposition depend 
on the parameters $a_1$...$a_K$, and there are only positive powers in the 
sum\footnote{ The negative powers of $\nu $ in the decomposition have no
practical use because we are interested in the quantity $G_\nu $ for large 
$\nu $.}.
  Note that function $G_\nu $ has a physical sense only for 
natural numbers $\nu $. However, the formulae (\ref{G}) allows us
to interpolate $G_\nu $ for any real number $\nu \geq 1$. Obviously,
one can imagine different interpolations but we shall use the simple
one\footnote{ Nevertheless, sometimes it is convenient to change a little 
this {\it canonical} interpolation. Indeed, the limit of $G_\nu $ when 
$\nu \to 1$ is equal to $c=\sum _{m=0}^Mc_m$, and nothing demands that 
it equals to zero (for example, see the drop model of spherical micelles, 
part 3).  Therefore, we can change the expression (\ref{G}) on the interval 
$(1,2)$ to obtain $\lim\limits _{\nu \to 1}G_\nu =0$. The simple way is to 
add function $-c/\nu ^x$ to $G_\nu $ where $x$ is a large positive number. 
We are interested in function $G_\nu $  for large $\nu $ thus the 
addition $-c/\nu ^x$ has no any influence on our results.} given by (\ref{G}). 
Furthermore, we shall differentiate some quantities by $\nu $, etc.

\vskip 1mm

  The important statement of the nucleation theory is following:
every minimum of aggregation work corresponds to a stable (or quasi-stable) 
state of the system of aggregates. We are interested in the formation of 
micelles, therefore there should be a ``deep'' minimum of $F_\nu $ 
(potential well). Moreover, this minimum is assumed to be unique and it will 
be denoted as $\nu _s$. Indeed, we are working with a small concentration of 
monomers, and consequently, there is only one stable state -- spherical micelles. 
Using the relation (\ref{F}), one can write the equation on $\nu _s$ as

\begin{equation}   \label{nu_s}
0=\frac{\partial F_\nu }{\partial \nu }(\nu _s)=
\frac{\partial G_\nu }{\partial \nu }(\nu _s)-\ln n_1 .
\end{equation}
Physically, the quantity $\nu _s$ is the aggregation number of
micelles, $\nu _s \gg 1$. In the practically important case the number
of amphiphiles 
in micelles is comparable (or more) to the concentration of monomers.

\vskip 1mm

  Using all previous assumptions we can establish our formalism
which defines the dependence of equilibrium concentration of monomers
on the parameters of micellizaton.

\vskip 10mm
\section{ Parametric equations of the theory of formation of
spherical micelles }

  Let us introduce a set of functions $f_\alpha (x)$ assuming $x>0$,

\begin{equation}   \label{f_alpha}
f_\alpha (x)=\sum\limits _{\nu =1}^{\infty }\nu ^{\alpha }e^{-G_\nu }x^{\nu }=
\sum\limits _{\nu =1}^{\infty }e^{-h(\nu )} ,
\end{equation}
where new function $h(\nu )$ is

\begin{equation}    \label{h}
h(\nu )=G_\nu -\nu \ln x-\alpha \ln \nu .
\end{equation}

  There are three important properties :

1) All functions $f_\alpha (x)$ are defined for any positive $x$ because
of a rapid growth of function $h(\nu )$ (since $G_\nu $ increases
faster than $\nu $). Moreover, $f_\alpha (x)\geq 0$ is true for any $x>0$.

\vskip 1mm
2) Simple verification shows that

\begin{equation}    \label{derives}
x\frac{\partial f_\alpha (x)}{\partial x}=f_{\alpha +1}(x).
\end{equation}
Hereof it follows that all $f_\alpha (x)$ are smooth
monotonously growing functions, with derivatives
of all orders.

\vskip 1mm
3) For $\alpha \ll \nu _s$ function $h(\nu )$ has a
minimum which is with high accuracy\footnote{
The equation for the minimum of $h(\nu )$ is derived by subtracting 
$\alpha /\nu $ in the right-hand side of the equation (\ref{nu_s}). 
Under condition $\alpha \ll \nu _s$ this term can be neglected by 
comparison with the positive powers of $\nu $.} equal to $\nu _s$.

\vskip 1mm

  Let us represent $f_\alpha (x)$ as a sum of three terms which
correspond to mono\-mers, small aggregates and micelles,

\begin{equation}   \label{sum}
f_\alpha (x)=x+\sum\limits _{\nu =2}^{\nu _0}e^{-h(\nu )}+\sum\limits _{\nu
=\nu _0}^{\infty }e^{-h(\nu )}=x+f^{0}_\alpha (x)+\tilde{f}_\alpha (x).
\end{equation}
It is necessary to say that the choice of this decomposition is artificial:
$\nu _0$ depends on the given model. Usually $\nu _0$ 
can be chosen as the maximum of $F_\nu $, i.e. a point where $n_\nu $ has
its minimum. As we take interest in formation of micelles, it is reasonable
to suppose that the contribution of micelles dominates over the
contributions of small aggregates,

\begin{equation}    \label{inequal}
f^{0}_\alpha (x)\ll \tilde{f}_\alpha (x).
\end{equation}
We neglect functions $f^{0}_\alpha (x)$, and the sum (\ref{sum}) becomes

\begin{equation}    \label{sum1}
f_\alpha (x)\approx x+\tilde{f}_\alpha (x)  .
\end{equation} 

\vskip 1mm

  Now we calculate an approximation to function $\tilde{f}_\alpha (x)$
(which corresponds to micelles) for small values of $\alpha $.
For this purpose we represent the corresponding sum in (\ref{sum})
as an integral, decompose $h(\nu )$ in the vicinity of its minimum
$\nu _s$ up to the square terms, and calculate the obtained gaussian integral
after replacing the lower integral limit by $-\infty $,

\begin{equation}   \label{ff}
\tilde{f}_\alpha (x)\approx \exp[-\beta (x)],  \hskip 5mm
\beta (x)=h(\nu _s)+\frac12\ln h''(\nu _s)-\frac12\ln (2\pi ) .
\end{equation}
Due to the independence of $\nu _s$ of $\alpha $, we obtain
an important relation between functions $\tilde{f}_\alpha $,

$$\tilde{f}_\alpha /\tilde{f}_\beta =(\nu _s)^{\alpha -\beta }, $$
whence we conclude

\begin{equation}   \label{relation}
\tilde{f}_\alpha =(\nu _s)^{\alpha -1}\tilde{f}_1  .
\end{equation}
In terms of functions $f_\alpha (x)$ the law of mass conservation
(\ref{masse}) can be written as equation for $n_1$,

\begin{equation}   \label{masse1}
f_1(x)=N.
\end{equation}
Due to the continuity and monotone growth of $f_\alpha (x)$, for any $N$ 
there is unique solution of (\ref{masse1}) which is equal to $n_1$,

\begin{equation}   \label{masse2}
f_1(n_1)=N.
\end{equation}
Substituting (\ref{ff}) and (\ref{masse2}) in (\ref{sum1}), we obtain
the transcendental equation for the equilibrium concentration of
monomers $n_1$ (cf.\cite{Mic4}),

$$N=n_1+\exp[-\beta (n_1)].$$
Practically this equation leads to the same difficulties in solution 
as the initial equation (\ref{masse2}). Therefore we continue to
develop our method.

\vskip 1mm

  Let us consider the equation (\ref{masse1}). Once this equation is 
solved, we shall find the dependence of $x$ (i.e. $n_1$) on the parameters
$a_1$...$a_K$. One can treat this dependence as a trajectory in 
the phase space based on parameters $a_1$...$a_K$, where function $f_1(x)$ 
is ``integral of motion'' (i.e. it is a constant on this trajectory). 
Fixing all parameters except one and differentiating (\ref{masse1}) by 
this single parameter $a_k$, we express the derivatives $dx/da_k$ according to

\begin{equation}    \label{dx_da}
\frac{dx}{da_k}=-\left[\frac{\partial f_1}{\partial x}\right]^{-1}\frac{\partial f_1}{\partial a_k} .
\end{equation}
The derivative in denominator can be found with the help of (\ref{derives}).
Now we calculate the derivative in nominator,

$$\frac{\partial f_1}{\partial a_k}=\sum\limits _{\nu =1}^{\infty }
\nu e^{-G_\nu }x^{\nu }\left(-\frac{\partial G_\nu }{\partial a_k}\right) .$$
Taking into account the representation of $G_\nu $ as a sum of fractional
powers of $\nu $, one has

\begin{equation}    \label{Cnk}
\frac{\partial G_\nu }{\partial a_k}=\sum\limits _mc_{m,k}\nu ^{\rho m}, \hskip 5mm
c_{m,k}=\frac{\partial c_m}{\partial a_k} .
\end{equation}
Collecting the previous expressions together, dividing by $x$ and
bringing it under the sign of differential, we obtain

\begin{equation}    \label{dlnx_da}
\frac{d\ln x}{da_k}=\frac{1}{f_2(x)}\sum\limits _mc_{m,k}f_{\rho m+1} .
\end{equation}

  Now we represent functions $f_\alpha $ according to (\ref{sum}), 
where all the terms $\tilde{f}_\alpha $ are expressed by (\ref{relation}) 
and all the terms $f^0_\alpha (x)$ are neglected by virtue of (\ref{inequal}),

$$\frac{d\ln x}{da_k}=\frac{1}{x+\nu _s\tilde{f}_1}
\left(x\sum\limits _mc_{m,k}+\tilde{f}_1\sum\limits _mc_{m,k}\nu _s^{\rho m}\right) .$$
In the first part we demanded  that the number of amphiphiles in micelles 
($\tilde{f}_1$) has the same or much greater order than in monomers ($n_1$),
therefore we can neglect\footnote{ Here we make second approximation, 
after the gaussian one in calculation of $\tilde{f}_\alpha $. The essential 
simplification brought by this operation is the independence of derivatives 
$d\ln x/da_k$ of $\tilde{f}_1$.}     $x$ and $x\sum _mc_{m,k}$ by comparison
with $\nu _s\tilde{f}_1$ and $\tilde{f}_1\sum _mc_{m,k}\nu _s^{\rho m}$
in nominator and denominator respectively. We have

\begin{equation}   \label{dlnx_da2}
\frac{d\ln x}{da_k}=\sum\limits _mc_{m,k}\nu _s^{\rho m-1} .
\end{equation}
With equation (\ref{nu_s}) we obtain a closed system of equations
for two unknown functions : $\ln x$ and $\nu _s$. Differentiating
the equation (\ref{nu_s}) by $a_k$,

$$\frac{d\ln x}{da_k}=\rho \sum\limits _mm\left(c_{m,k}\nu _s^{\rho m-1}+
c_m(\rho m-1)\nu _s^{\rho m-2}\frac{d\nu _s}{da_k}\right) , $$
we establish one differential equation for $\nu _s$. After simplification,
this equation gets a form

\begin{equation}   \label{eqn}
\frac{d\nu _s}{da_k}=-\frac{\sum\limits _mc_{m,k}(\rho m-1)\nu _s^{\rho m}}
{\sum\limits _mc_m\rho m(\rho m-1)\nu _s^{\rho m-1}} , \hskip 10mm (k=1..K).
\end{equation}
  So, we have obtained a set of equations which define the dependence of
the aggregation number of micelles $\nu _s$ on the parameters of the theory. 
We shall call these equations {\it parametric}.
Its solution is connected with the equilibrium concentration of
monomers $n_1$ via (\ref{nu_s}).

\vskip 1mm

  If we are interested only in dependence on one specific parameter $a_k$, 
it is sufficient to solve one of the equations (\ref{eqn}) for this 
parameter. However, in this case there appears an arbitrary function of 
the other parameters. On the contrary, if we search for a complete solution,
it is necessary to solve all the equations (\ref{eqn}) and to accommodate
all appearing arbitrary functions. Note that in the case of a large
number of parameters this algorithm can be very complicated.

\vskip 1mm

  We make another important remark. If one adds an arbitrary function
$g(\nu )$ to the expression of $G_\nu $ which {\it does not depend}
on the set of parameters $a_1$...$a_K$ and {\it does not break a convergence}
of the series (\ref{f_alpha}), then the equation (\ref{eqn}) does not
change. Indeed, function $g(\nu )$ disappears after differentiation
by the parameters $a_1$..$a_K$. Consequently, $g(\nu )$ has no
influence\footnote{ It is more accurate to say that function $g(\nu )$ 
has no influence on the analytic form of solution, but it changes, of course,
initial conditions, and consequently, some constants of solution.} 
on a solution of the parametric equations (\ref{eqn}).
On the contrary, the addition of $g(\nu )$ affects the equation
(\ref{nu_s}), and the expression of equilibrium concentration of
monomers will change according to the formula

\begin{equation}    \label{n'}
n'_1=n_1\exp [-g'(\nu _s)],
\end{equation}
where $n_1$ is a solution of (\ref{eqn}) without $g(\nu )$. This
simple proposition is very useful from the practical point of view.
We see that the characteristic $\nu _s$ is universal, its equation
(\ref{eqn}) is independent of a small variations of function $G_\nu $.
On the contrary, the equilibrium concentrations of monomers and micelles
depend on the derivative of $g(\nu )$ according to (\ref{n'}).

\vskip 1mm

  The common property of the set of equations (\ref{eqn}) is their 
independence of the linear term $\nu $ which disappears by virtue of 
a factor $(\rho m-1)$ when $m$ equals to $m_0=1/\rho $ corresponding
to a linear term (in the case when $1/\rho $ is not natural number,
a priori there is no linear term). Nevertheless, sometimes\footnote{
See, for example, Grinin's model, part 4.} there is a dependence 
of the equation (\ref{eqn}) on the coefficient $c_{m_0}$ by virtue of 
the term $c_0$ which can contain this coefficient. However, usually one 
can neglect the term $c_0$ by comparison with $\nu _s^{\rho m}$ 
($\nu _s\gg 1$). Note that every case should be analyzed particularly.

\vskip 1mm

  What is the qualitative information that is contained in the
equation (\ref{eqn})? It seems possible to use this equation
(and the whole method) to verify the following important hypothesis 
in the theory of micellization. As it was mentioned above,
the exact expression for the aggregation work 
is unknown, therefore we use the model expressions (see, for example,
(\ref{Ga})). However, it is natural to suppose that all physically 
measured quantities depend on global characteristics of the 
aggregation work (such as minimum of $F_\nu $, depth of this minimum,
height of activation barrier, etc), rather than on a choice of
its particular expression.
If this hypothesis works, it is sufficient to take a simple expression
for $F_\nu $ satisfying to the qualitative requests and to solve
the problems with the use of this expression.

\vskip 10mm
\section{ Solution of the parametric equations for the drop model
of spherical micelles.}

  Let us consider a {\it drop model} of spherical micelles,
firstly elaborated by Is\-raelachvili, Mitchell and Niham (see \cite{Israel}) 
and reformulated independently by Rusanov in terms of the aggregation 
work (see \cite{Rus}). The basic ideas of this model are simple :

1) hydrophilic electrically charged heads form a surface of micelle 
(sphere);

2) hydrophobic tails are very flexible, and they form a liquid drop 
in the micellar core. 

There are three terms in the aggregation work : 

 -- bulk term (proportional to the number $\nu $ of amphiphiles in
the aggregate);
 
 -- surface term corresponding to the attractive hydrophobic forces
(and proportional to $s\nu $, where $s$ is the surface area per amphiphile ;
$s\nu \sim \nu ^{2/3}$ for the spherical micelles) ;

 -- surface term corresponding to the repulsive electrostatic forces
(proportional to $\nu /s\sim \nu ^{4/3}$ as it has been shown by 
Tanford, \cite{Tanford}).

Leaving out the details, we write only the final expression for
the aggregation work for this model,

\begin{equation}   \label{G_Rus}
G_\nu =b_1\nu ^{4/3}-\left(\frac43\sqrt{2b_1b_3}\right)\nu +b_3\nu ^{2/3},
\end{equation}
where the parameters\footnote{ We use Rusanov's notations.}
$b_1$ and $b_3$ are defined through the physical characteristics,

\begin{equation}   \label{b_1_3}
b_1=\frac{(ez)^2\delta }{8\pi \ve _0\ve \lambda ^2kT} , \hskip 10mm
b_3=\frac{4\pi \lambda ^2\gamma _0}{kT} ,
\end{equation}
where $\lambda =(3v/4\pi )^{1/3}$ is radius of a sphere which has
the same volume as one hydrocarbon chain, $ez$ is electric charge of 
one amphiphile,  $\ve _0$ is dielectric constant, $\ve $ is 
dielectric permittivity, $\delta $ is separation of the capacitor planes, 
$\gamma _0$ is surface tension of the hydrocarbon-water interface.

  The expression (\ref{G_Rus}) defines the coefficients of decomposition
of the aggregation work by the powers of $\nu $ with $\rho =1/3$,

$$c_0=0, \hskip 4mm   c_1=0, \hskip 4mm   c_2=b_3, \hskip 4mm  
c_3=-\frac43\sqrt{2b_1b_3}, \hskip 4mm  c_4=b_1 .$$

  In the framework of this model the parametric equations (\ref{eqn})
can be solved exactly, without any approximation (however, the parametric 
equations themselves remain approximate ones). Substituting the
values of $c_m$ into the equations (\ref{eqn}), we obtain

\begin{eqnarray}    \label{eqn_Rus}
\frac{d\nu _s}{db_1} &=& -\frac32\frac{\nu _s^{4/3}}{2b_1\nu _s^{1/3}-b_3\nu _s^{-1/3}} ,\\
\frac{d\nu _s}{db_3} &=&  \frac32\frac{\nu _s^{2/3}}{2b_1\nu _s^{1/3}-b_3\nu _s^{-1/3}} .
\end{eqnarray}
The substitution

$$\nu _s=(b_3/b_1)^{3/2}\varphi ^3(b_1,b_3)$$
leads to

$$2b_1\frac{d\varphi }{db_1}=\frac{\varphi (\varphi ^2-1)}{2\varphi ^2-1} , \hskip 10mm
-b_3\frac{d\varphi }{db_3}=\frac{\varphi (\varphi ^2-1)}{2\varphi ^2-1} , $$
and we obtain the solution

\begin{eqnarray}   \label{y2}
\varphi ^2(\varphi ^2-1) &=& \gamma _3(b_3)b_1 ,\\
\varphi ^2(\varphi ^2-1) &=& \gamma _1(b_1)b_3^{-2} .
\end{eqnarray}
Equaling the right-hand sides, we relate functions $\gamma _1(b_1)$ and
$\gamma _3(b_3)$,

$$\gamma _1(b_1)b_1^{-1}=\gamma _3(b_3)b_3^2. $$
As the parameters $b_1$ and $b_3$ are independent, both sides of
this equality are constant, whence

$$\gamma _1(b_1)=\gamma (N)b_1,   \hskip 5mm
\gamma _3(b_3)=\gamma (N)b_3^{-2},$$
where a constant $\gamma (N)$ can be determined numerically
for any $N$ (see table 1).
  Substituting function $\gamma _1(b_1)$ in the equation (\ref{y2}),
one finds the dependence of the aggregation number of micelles $\nu _s$ 
on the parameters of micellization,

\begin{equation}   \label{nus_Rus}
\nu _s=(\alpha b_3/2b_1)^{3/2},  \hskip 10mm
\alpha =1+\sqrt{1+4\gamma (N)b_1b_3^{-2}}  .
\end{equation}
  The equilibrium concentration of monomers can be found
by substitution of (\ref{nus_Rus}) in the equation (\ref{nu_s}),

\begin{equation}   \label{n1_Rus}
n_1=\exp\left(\frac23\sqrt{2b_1b_3}\left[\alpha ^{1/4}-\alpha ^{-1/4}\right]^2
\right)   .
\end{equation}

\vskip 1mm

  Now it is useful to evaluate the values of parameters $b_1$ and $b_3$.
We can obtain\footnote{ Note that these estimations are quite enough for 
our purposes. Indeed, we are working with the simple model, therefore we 
can not expect a high accuracy of predicted results. }

\begin{equation}  \label{b3}
b_1\approx 2, \hskip 10mm  b_3\approx 30 
\end{equation}
for sodium dodecyl sulfate ($C_{12}H_{25}OSO_3Na$) in water at the ambient
temperature ($T=300 K$) using characteristic values of the following
parameters (see \cite{Israel2}, \cite{Israel}) :

 -- volume of the hydrocarbon tail of an amphiphile, $v=0.350 \: nm^3$ ; 

 -- interfacial free energy per unit area of aggregate, 
$\gamma _0=50 \: erg/cm^2$ ;

 -- optimal surface area per amphiphile, $a_0\approx 0.62 \: nm^2$
(this value was used to calculate $b_1$, see for details \cite{Israel2}).

\vskip 1mm

  We have determined the dependence of equilibrium concentration of mono\-mers 
on the parameters $b_1$ and $b_3$. The dependence on the total concentration 
of amphiphiles $N$ is contained in the constant $\gamma (N)$ which can be 
numerically calculated for any $N$. Indeed, it is sufficient to find $n_1^0$ 
for one set of $b_1^0$, $b_3^0$ and $N$ by solving the equation (\ref{masse}) 
or (\ref{masse2}), then to substitute the value of $n_1^0$ in the
expression (\ref{n1_Rus}) and to express the constant $\gamma $.
With the help of numerical simulations we obtain the values of $\gamma (N)$ 
for some $N$ represented in the table 1.

\begin{center}
\begin{tabular}{| c | c | c | c | c | c | c | c |}  \hline
$N$ & 2.0 & 2.5 & 3.0 & 3.5 & 4.0 & 4.5 & 5.0 \\  \hline
$n_1$ & 1.9865 & 2.1159 & 2.1471 & 2.1648 & 2.1771 & 2.1865 & 2.1942 \\  \hline
$\gamma $ & -32.774 & -22.852 & -20.486 & -19.156 & -18.232 & -17.525 & -16.953 \\  \hline
\end{tabular}
\end{center}
{\em Table 1. The values of constant $\gamma (N)$ and of equilibrium
concentration of monomers $n_1$ as functions of $N$. For these
simulations we have used $b_1=2.0$, $b_3=30$ corresponding to sodium
dodecyl sulfate in water. }
\vskip 2mm

  Note that the aggregation work $F_\nu $ for this model 
grows rapidly for small $\nu $ ; thus dimers, trimers and 
other small aggregates are completely negligible. Consequently, 
the number of amphiphiles in micelles can be well approximated
as $N-n_1$. 

\vskip 1mm

  The essential advantage of the expression (\ref{n1_Rus}) is its
explicit form which allows to analyze dependencies of all the
equilibrium concentrations on parameters $b_1$ and $b_3$. Moreover, 
this formula has a quite high accuracy. The table 2 represents
its relative errors (in per cent) when parameters $b_1$ and
$b_3$ are widely varied. 

\begin{center}
\begin{tabular}{| c | c | c | c | c | c | c | c | c | c | c | c | c |}  \hline
$b_3 \backslash b_1$ & 1.0 & 1.5 & 1.6 & 1.7 & 1.8 & 1.9 & 2.0 & 2.1 & 2.2 & 2.3 & 2.4 & 2.5 \\ \hline
28 & 0.9 & 0.5 & 0.3 & 0.1 & 0.0 & 0.3 & 0.5 & 0.7 & 1.0 & {\it 1.3} & {\it 1.6} & {\it 1.9} \\
29 & 0.9 & 0.6 & 0.4 & 0.3 & 0.1 & 0.0 & 0.2 & 0.4 & 0.7 & 0.9 & {\it 1.2} & {\it 1.5} \\
30 & 0.9 & 0.7 & 0.6 & 0.4 & 0.3 & 0.2 &  0  & 0.2 & 0.4 & 0.9 & 0.8 & {\it 1.0} \\
31 & 0.9 & 0.7 & 0.7 & 0.6 & 0.5 & 0.3 & 0.2 & 0.1 & 0.1 & 0.6 & 0.4 & 0.6 \\
32 & 0.9 & 0.8 & 0.7 & 0.7 & 0.6 & 0.5 & 0.4 & 0.3 & 0.2 & 0.0 & 0.1 & 0.2 \\ \hline
\end{tabular}
\end{center} 
{\em Table 2. The relative errors (in per cent) of the formula
(\ref{n1_Rus}) when $b_1$ changes from $1.0$ to $2.5$, $b_3$ changes
from $28$ to $32$. Generally, these errors are less than $1\%$
(beside of the right upper corner marked in italic that is 
not physical, see below). } 
\vskip 2mm

  Note that the variation of $b_3$ from $30$ to $32$ ($28$)
can be caused by decreasing (increasing) the temperature by $20^\circ C$
(i.e. the domain of variation of temperature is between $5^\circ C$ 
and $45^\circ C$). On the contrary, the parameter $b_1$ can be varied
in the wide enough region. Indeed, if one adds salts (for example, $NaCl$) 
in solution, they will screen electrostatic repulsion of amphiphile heads,
i.e. the parameter $b_1$ will be decreased. Normally, it is
difficult to increase $b_1$ considerably, thus the right upper
corner of the table 2 is not physical. We can conclude that the
formula (\ref{n1_Rus}) approximates the equilibrium concentration
of monomers with relative errors less than $1\%$.

\newpage
\section{ Solution of the parametric equations for Grinin's model of
spherical micelles.}

   There is another model of spherical micelles elaborated by Grinin
\cite{Grinin}. The hydrophobic tails of amphiphiles are assumed to be 
enough ordered in the core of micelles (on the contrary, for the drop 
model we assumed that hydrophobic tails are completely flexible as a
liquid). The geometric restrains imply that the hydrophobic contribution
to chemical potential of one amphiphile in aggregate is proportional to
$\nu ^{1/2}$. The hydrophilic effects can be described by the term 
proportional to $\nu $ (see \cite{Rus2}). The more detailed description 
can be found in the original article \cite{Grinin}. 

  For our purposes the final expression for the aggregation work is important,

\begin{equation}    \label{Ga}
G_\nu =\frac{b}{2}(\nu^2-1)-\frac{2a}{3}(\nu ^{3/2}-1)+\frac{a^2}{4b}(\nu -1),
\end{equation}
the two parameters\footnote{ As for the drop model, the parameters
$a$ and $b$ are not physical, but they are connected with real
physical characteristics,

$$a=B\frac{d+d_{H_2O}-s^{1/2}}{(4\pi )^{1/2}l_c},  \hskip 5mm
b=\frac{(ze)^2}{4\pi \ve _0\ve l_ckT}\left[(n_c+r/l_c)^{-1}-
(n_c+r/l_c+\delta /l_c)^{-1}\right] , $$
where $B$ is empiric constant (which equals to $1.4$ under normal
conditions, see \cite{Rus}), $d$ is minimal distance between
hydrophobic tail and water molecule, $d_{H_2O}$ is characteristic
size of water molecule, $l_c$ is hydrocarbon length, $s$ is 
effective surface area of hydrophobic tail, $\delta $ is characteristic
scale of space distribution of charge formed by aggregate, $ez$ is
effective charge of hydrophilic head of amphiphile, $\ve _0$ is
dielectric constant, $\ve $ is dielectric permittivity of solution,
$n_c$ is number of carbon atoms in hydrophobic tail.}
are : $a_1=a$ and $a_2=b$.
The expression (\ref{Ga}) is the decomposition of the aggregation work 
by the powers of $\nu $ with $\rho =1/2$ and with the coefficients

$$c_0=\frac{2a}{3}-\frac{b}{2}-\frac{a^2}{4b}, \hskip 5mm
c_1=0, \hskip 5mm  c_2=\frac{a^2}{4b}, \hskip 5mm c_3=-\frac{2a}{3},
\hskip 5mm  c_4=\frac{b}{2}. $$
After calculating matrix of derivatives $c_{m,k}$ and substituting it in
(\ref{eqn}), we obtain two equations,

\begin{eqnarray}   \label{da}
\frac{d\nu _s}{da} &=& \frac23\frac{\nu _s^{3/2}}
{2b\nu _s-a\nu _s^{1/2}} , \\  \label{db}
\frac{d\nu _s}{db} &=& -\frac{\nu ^2_s}{2b\nu _s-a\nu _s^{1/2}}
\end{eqnarray}
(in the nominator we have neglected a small term $(a/2b-2/3)$ by
comparison with $\nu _s^{3/2}$ and $\nu _s^2$). The substitution

\begin{equation}   \label{varphi2}
\nu _s =(a/b)^2\varphi ^2(a,b)
\end{equation}
reduces these equations to the quadratures

$$-\frac{2da}{a}=\frac{d\varphi (2\varphi -1)}{\varphi (\varphi -2/3)} , \hskip 10mm
\frac{3db}{2b}=\frac{d\varphi (2\varphi -1)}{\varphi (\varphi -2/3)} ,$$
which give after integration two equations of fourth degree connecting
$\varphi $ with $a$ and $b$,

\begin{eqnarray*}
\varphi ^3|\varphi -2/3| &=& \gamma _1(b)a^{-4} , \\
\varphi ^3|\varphi -2/3| &=& \gamma _2(a)b^3 ,
\end{eqnarray*}
where $\gamma _1(b)$, $\gamma _2(a)$ are arbitrary functions.
Equaling the right-hand sides, one obtains : $\gamma _1(b)=Cb^3$,
$\gamma _2(a)=Ca^{-4}$, i.e. the equation is

\begin{equation}   \label{varphi0}
\varphi ^3|\varphi -2/3|=Ca^{-4}b^3 ,
\end{equation}
with an arbitrary constant $C$.
Using small variations from the initial parameters, one can conclude 
that the factor $\varphi ^3$ changes a little\footnote{
Indeed, using an explicit expression (\ref{Ga}) for function $G_\nu $
we obtain an equation for extrema,
$$b\nu -a\nu ^{1/2}+a^2/4b=\ln n_1,  $$
with larger solution (corresponding to the minimum $\nu _s$)

$$\nu _s=\bigl(a/2b+\sqrt{(\ln n_1)/b}\bigr)^2 ,$$
whence it follows that $\varphi \geq 1/2$, therefore all changes
occur in the vicinity of the point $2/3$.}. Then we obtain an
approximate solution

\begin{equation}    \label{varphi}
\varphi =2/3+(Ca_0^4b_0^{-3})a^{-4}b^3,
\end{equation}
or, with the help of (\ref{varphi2}), we have

\begin{equation}   \label{nu_sa}
\nu _s=(2a/3b+\gamma _G(N)a^{-3}b^2)^2,
\end{equation}
where a constant $\gamma _G$ can be defined for any $N$,
and once defined (by the numerical solution of initial
equation (\ref{masse}) for certain fixed parameters $a_0$, $b_0$),
it becomes universal characteristic for the model under consideration. 
In the table 3 we show the values of $\gamma _G(N)$ for some $N$.
  Now one can use the equation (\ref{nu_s}) to calculate $n_1$,

\begin{equation}   \label{n1}
n_1=\exp\left[\frac{a^2}{36b}(1+\gamma _G(N)a^{-4}b^3)^2\right].
\end{equation}
Thus we have found the complete dependence of the equilibrium
concentration of monomers on the parameters for Grinin's model
of spherical micelles.

\vskip 1mm

  Note that the characteristic values of the parameters $a$ 
and $b$ of Grinin's model of spherical micelles are

\begin{equation}   \label{a_b}
a\approx 1.5,  \hskip 5mm  b\approx 0.1 .
\end{equation}
With the help of numerical simulations, we calculate the values of 
the constant $\gamma _G$ for some $N$ for these characteristic
values of the parameters $a$ and $b$, see table 3.

\newpage 

\begin{center}
\begin{tabular}{| c | c | c | c | c | c | c | c |}  \hline
$N$ &  2.0 & 2.5 & 3.0 & 3.5 & 4.0 & 4.5 & 5.0 \\ \hline
$n_1$ & 1.6280 & 1.6460 & 1.6549 & 1.6608 & 1.6652 & 1.6687 & 1.6716  \\  \hline
$\gamma _G$ & -592.08 & -541.82 & -517.62 & -501.66 & -489.76 & -480.29 & -472.42 \\ \hline
\end{tabular}
\end{center}
{\em Table 3. Grinin's model: values of the constant $\gamma _G$
and of the equilibrium concentration of monomers $n_1$
for some $N$ calculated with the parameters $a=1.5$, $b=0.1$.}
\vskip 2mm

  The table 4 contains\footnote{ To fill this table,
the equilibrium concentration of monomers $n_1$ for some values
of parameters $a$ and $b$ was calculated by two different methods :
by numerical solution of the equation (\ref{masse2}) and with
the help of the approximate formula (\ref{n1}).} the relative errors
(in per cent) of the expression (\ref{n1}).

\begin{center}
\begin{tabular}{| c | c | c | c | c | c | c | c | c | c | c | c |}  \hline
 $b\hskip 2mm \backslash \hskip 2mm a$ & 1.0 & 1.1 & 1.2 & 1.3 & 1.4 & 1.5 & 1.6 & 1.7 & 1.8 & 1.9 & 2.0 \\  \hline
0.05 & 0.0 & 0.9 & --  & --  & --  & --  & --  & --  & --  & --  & --  \\
0.06 & 0.7 & 0.0 & 0.8 & --  & --  & --  & --  & --  & --  & --  & --  \\
0.07 & 1.2 & 0.8 & 0.1 & 0.8 & --  & --  & --  & --  & --  & --  & --  \\
0.08 & 0.9 & 1.3 & 0.9 & 0.1 & 0.8 & --  & --  & --  & --  & --  & --  \\
0.09 & 1.3 & 1.2 & 1.4 & 0.9 & 0.0 & 0.9 & --  & --  & --  & --  & --  \\
0.10 & --  & 0.2 & 1.4 & 1.4 & 0.8 &  0  & 1.0 & --  & --  & --  & --  \\
0.11 & --  & --  & 0.6 & 1.6 & 1.4 & 0.7 & 0.1 & 1.2 & --  & --  & --  \\
0.12 & --  & --  & --  & 1.1 & 1.5 & 1.3 & 0.6 & 0.3 & 1.5 & --  & --  \\
0.13 & --  & --  & --  & 0.4 & 1.5 & 1.7 & 1.3 & 0.5 & 0.5 & 1.8 & --  \\
0.14 & --  & --  & --  & --  & 0.5 & 1.7 & 1.7 & 1.1 & 0.3 & 0.7 & 2.2 \\
0.15 & --  & --  & --  & --  & --  & 1.1 & 1.8 & 1.6 & 1.0 & 0.0 & 1.0 \\  \hline
\end{tabular}
\end{center}
{\em Table 4. Relative errors (in per cent) of the expression (\ref{n1})
calculated for a wide domain of the parameters $a$ and $b$ ($N=2$).}
\vskip 2mm

  The absence of the values in the left low corner is caused by the
condition $n_1>1$ (otherwise there are no stable aggregates). The right
upper corner is empty by virtue of the condition $n_1<N$ (the opposite
case has no physical sense). We see that the relative error of the
approximate solution (\ref{n1}) changes in an irregular way,
however, it does not exceed $2\% $ when the parameters $a$ and $b$ 
take all physically reasonable values. This accuracy can be considered 
as a high one, thus we conclude that the formula (\ref{n1}) is a good 
approximation for the equilibrium concentration of monomers in Grinin's model. 
Nevertheless, there are problems which demand more accurate results. For this 
reason we would like to stress that the form of our expressions is explicit, 
therefore it is possible to analyze the results qualitatively and to compare 
different models.

\vskip 1mm

  Note that the form of the solution (\ref{n1}) is rather
different from drop's one (\ref{n1_Rus}). However, we can not compare 
them directly because the parameters $a$, $b$ and $b_1$, $b_3$
are completely incompatibles. Thus, one should express these parameters
in terms of the physical quantities (such as the global characteristics 
of the aggregation work), substitute their into (\ref{n1_Rus}) and (\ref{n1}),
and after that to compare the results.

\vskip 10mm
\section{ Conclusions }

  Based on the nucleation theory, we have developed the method
which allows to obtain the parametric equations under rather
general assumptions :

-- closure of the system ;

-- dilute solution of surfactants ;

-- general model expression of $G_\nu $ as a sum of
the fractional powers of $\nu $ ;

-- comparability of the numbers of amphiphiles in monomers and micelles.

To deduce the parametric equations we have used these assumptions,
but they are of the rather different significance.
  More important hypothesis is the closure of system.
Some other ones are necessary to be mathematically
rigorous. Finally, there are certain conditions which are used
to specify the differential equations, to obtain concrete results 
(such as the solutions (\ref{n1_Rus}) or (\ref{n1})),
but they are not strictly necessary for the method's work.
For example, the decomposition of function $G_\nu $ by
the fractional powers of $\nu $ is physically reasonable and
it simplifies the method significantly. But this assumption
can be diminished, i.e. the expression of $G_\nu $ can
be generalized. In other words, by concretizing the form of some
expressions (for example, (\ref{F}), (\ref{G}), etc.),
we simplify mathematically our problem, physically leaving it
in general form.

\vskip 1mm

  We have obtained the differential equation for the important
characteristic of micellization -- for the aggregation number of micelles. 
Taking into account the remark in second part about the invariance
of the parametric equations by addition of some function $g(\nu )$
to the aggregation work, we can conclude that
$\nu _s$ is an universal quantity of micellization. On the
contrary, the equilibrium concentrations of monomers and micelles
essentially depend on the concrete expression $G_\nu $, consequently,
they are not universal.

  The system of parametric equations was solved for two
practically important model of spherical micelles :
the drop and Grinin's models. We have obtained an explicit
dependence of the equilibrium concentration of monomers
on the parameters of micellization. The numerical simulations
showed that these formulae have a high accuracy.

\vskip 1mm

  At last, there appears a real possibility to verify
the hypothesis that the physically measured quantities of 
micellization (e.g., the relaxation times) are independent of the 
{\it concrete} analytical expression of the aggregation work, but 
they are defined by its global characteristic (such as point of minimum, 
height of activation barrier, etc). If this hypothesis will be confirmed,
it will justify the use of the simple model expressions of the 
aggregation work (such as (\ref{G_Rus}) for the drop model, or (\ref{Ga}) 
for Grinin's model). Otherwise, our method allows to compare different 
model expressions and the solutions based on them, and to determine the 
substantiality of the model and its accuracy.

\vskip 2mm

  We can conclude that elaborated method is effective enough both
for applications (for example, to determine the dependencies
of equilibrium concentration of monomers on the parameters for
the two models of spherical micelles) and for the general analysis.
We hope that it will allow to solve some problems in the theory of micelles,
and to comprehend profoundly the mechanism of micellization.

\vskip 10mm
\section*{ Acknowledgement  }

  The author would like to thank Professor A.P.Grinin whose 
invaluable advice helped to develop the present method.

\end{document}